\documentclass{aa}
\usepackage{graphicx}
\usepackage{natbib}
\usepackage{txfonts}
\usepackage{epstopdf}
\usepackage[utf8]{inputenc}
\usepackage[english]{babel}
 
\bibpunct{(}{)}{;}{a}{}{,} 
\usepackage{xcolor}

\def\apjs{ApJS}
  
\def\jgr{J. Geophys. Res.}

\def\aap{Astron. Astrophys.}

\def\apj{Astrophys. J.}

\def\spsr{Space Sci Rev}

\def\atcp{Atmos. Chem. Phys.}

\def\lrsph{Liv. Rev. Sol. Phys.}

\def\araap{Annu. Rev. Astron. Astrophys.}

\begin{document}

\title{Solar-cycle irradiance variations over the last four billion years}
\author{Anna V. Shapiro
          \inst{1}
          \and
          Alexander I. Shapiro\inst{1}
          \and
          Laurent Gizon\inst{1,2,3}
          \and
          Natalie A. Krivova\inst{1} 
          \and
          Sami K. Solanki\inst{1,4} 
          }

   \institute{Max-Planck-Institut f\"ur Sonnensystemforschung, Justus-von-Liebig-Weg 3, 37077 G\"ottingen\\         
   \email{shapiro@mps.mpg.de}
         \and
             Institut f\"ur Astrophysik, Georg-August-Universit\"at, Friedrich-Hund-Platz 1, 37077 G\"ottingen 
          \and
             Center for Space Science, NYUAD Institute, New York University Abu Dhabi, Abu Dhabi, UAE 
          \and
             School of Space Research, Kyung Hee University, Yongin, 446-701 Gyeonggi, Korea 
}

   \date{Received  ; accepted }

 
  \abstract
   {The variability of the spectral solar irradiance (SSI) over the course of the 11-year solar cycle is one of the manifestations of solar magnetic activity. There is a strong evidence that the SSI variability has an effect on the Earth's atmosphere. The faster rotation of the Sun in the past lead to a more vigorous  action of solar dynamo and thus potentially to larger amplitude of the SSI variability on the timescale of the solar activity cycle. This could led to a stronger response of the Earth's atmosphere as well as other solar system planets' atmospheres to the solar activity cycle. 
   }
   {We calculate the amplitude of the SSI and TSI variability over the course of the solar activity cycle as a function of solar age.}
   { We employ the relationship between the stellar magnetic activity and the age based on observations of solar twins. Using this relation we reconstruct solar magnetic activity and the corresponding solar disk area coverages by magnetic features (i.e. spots and faculae) over the last four billion years. These disk coverages are then used to calculate the amplitude of the solar-cycle SSI variability as a function of wavelength and solar age.} 
   {Our calculations show that the young Sun was significantly more variable than the present Sun. The amplitude of the solar-cycle Total Solar Irradiance (TSI) variability of the 600 Myr old Sun was about 10 times larger than that of the present Sun. Furthermore, the variability of the young Sun was spot-dominated (the Sun being brighter at the activity minimum than in the maximum), i.e. the Sun was overall brighter at activity minima than at maxima. The amplitude of the TSI variability decreased with solar age until it reached a minimum value at 2.8 Gyr. After this point, the TSI variability is faculae-dominated (the Sun is brighter at the activity maximum) and its amplitude increases with age.}
   {}



   {}


   \maketitle
%

\section{Introduction}
The action of a hydromagnetic dynamo generates magnetic field in the solar interior \citep{Charbonneau2010,Brun_LR}. This field becomes unstable, rises through the convective zone and finally emerges on the solar surface, leading to various manifestations of solar magnetic activity \citep{Sami_B}.  In particular, concentrations of the magnetic field on the solar surface form magnetic features, such as bright faculae and dark spots. The facular and spot contributions to the solar brightness usually do not fully compensate each other and the resulting imbalance defines the active (i.e. magnetic) component of the solar brightness variability.

As stars evolve on the main sequence (MS) they gradually spin down due to the magnetic braking \citep{Guebel2007, Reiners2012, Metcalfe2016}. 
Slowing of the stellar rotation weakens the stellar dynamo and, consequently, stellar magnetic activity and its contribution to the stellar brightness variations 
\citep[see e.g.][]{Noyes1984}. Such changes of the stellar brightness caused by the spin down have been observed. For example, \cite{Guinan2003} investigated the evolution of the 920--1180 {\AA} flux by studying six G0–G5 solar analogs with different ages. Using the selected stars as proxies for the younger Sun, they concluded that the 920--1180 {\AA} solar flux in the past was higher than the present-day level. It was 2 times higher 2.5 Gyr ago and 4 times higher 3.5 Gyr ago.


In addition to the gradual decrease of the magnetic activity with the solar age it also fluctuates on various timescales \citep{Balogh2014}. The most prominent of them is the timescale of the solar activity cycle which corresponds to the oscillation between toroidal and poloidal components of the solar magnetic field \citep{Parker1955}. The present duration of the solar activity cycle is  on average 11 years but there is evidence that it was shorter for the younger Sun \citep{Galanza2019}.

 The 11-year activity cycle is clearly visible in the space-borne measurements of the Total (TSI) and Spectral Solar Irradiance (SSI) \citep[see e.g. reviews by][]{Solanki2013, Ermolli2013}. The present-day TSI change due to the 11-year solar cycle is about 0.1 \%, while the variability in the UV spectral domain is significantly larger. Thus, it reaches e.g. several percent at 200-250 nm spectral range and tends to increase with decreasing wavelength \citep{Floyd2003}. There have been numerous studies indicating that the Earth's atmosphere responds to the 11-year cycle in SSI \citep [see e.g.][]{Meehl2009, Shapiro2013, Hood2015, Maycock2018}.
 
 The stellar activity cycles are also visible in the records of emission in the cores of the \ion{Ca} \relax II H and K lines \citep{Wilson1978}. The emission in these lines is strongly modulated by the magnetic heating of the chromosphere and is thus often used for quantitative characterization of stellar magnetic activity \citep{Baliunas1995}. It was found that the amplitude of the activity cycle in  \ion{Ca} \relax II H and K emission depends on the mean (i.e. averaged over the activity cycle) level of magnetic activity: more active stars have more vigorous activity cycles \citep{halletal2009, Egeland_PHD, Radick2018}. This implies that the amplitude of the solar activity cycle was stronger for the younger Sun than it is now. This could lead to a larger irradiance variability and, consequently, to a larger impact of the irradiance variability on the Earth's atmosphere.

 In this study we reconstruct the amplitude of the SSI variability as a function of wavelength and solar age. For this we employ the model of stellar variability developed by \cite{Shapiro2014}, which is based on the SATIRE approach \citep[][]{Krivova2003}. Furthermore,  we use the relation between magnetic activity and age recently established by \cite{LorenzoOliveira2018} for a sample of solar twins.  We organize this paper as follows. Sect.~2 describes the activity-age relation, Sect.~3  presents the model, the results are presented in  Sect.~4. Conclusions are summarized in Sect.~5.

\section{Scenarios of evolution of solar activity}
The variability of solar irradiance is driven by the change of the solar surface coverages by active features. In turn, the coverages strongly depend on the solar activity. Thus the first step in modeling cyclic SSI variability to the past is to determine the dependence of solar magnetic activity on age. More specifically, for the purposes of this study we need to reconstruct the S-index, which is proportional to the ratio between the flux in \ion{Ca} \relax II H and K emission lines and two nearby pseudo-continuum bands \citep[see, e.g.][]{radicketal1998}. The reconstructed S-index will then be used as an input into the model described in Sect.~3 to obtain the SSI variability.

Several different stellar activity-age relations for the MS stars have been proposed \citep [see e.g.][]{Lachaume1999, Mamajek2008, LorenzoOliveira2018}. One of the main uncertainties in defining this relation lies in estimation stellar age \citep [see e.g. review by] [] {Soderblom2010}. Such estimations, done with stellar evolutionary models, are complicated by poorly known values of luminosity. To overcome this \cite{LorenzoOliveira2018} have considered a sample of solar twins with near-solar value of luminosity. They defined solar twins as  stars with $T_{\rm eff}$ within $\pm$100K of the solar value, log g within $\pm$0.1 dex and [Fe/H] within $\pm$0.1 dex of the solar values. This allowed them to determine the ages more accurately, leading to the relationship: 
\begin{equation}
 \log t = 0.0534 - 1.92 \log R'_{\rm HK},
\end{equation}
where $t$ is stellar age in Gyr and log is the base-ten logarithm.

 The activity here is defined via the dimensionless  $\log R'_{\rm HK}$ parameter, which is a function of the S-index and the stellar colour B-V \citep{radicketal1998}. In contrast to the S-index, the $R'_{\rm HK}$ index does not include the photospheric contribution to the emission in the \ion{Ca} \relax II H and K lines. This has to be taken into account when converting the $R'_{\rm HK}$ value into the S-index. Consequently, we add photospheric contribution to $R'_{\rm HK}$ and calculate $R_{\rm HK}$ value:
\begin{equation}
 R_{\rm HK} = R_{\rm phot} + R'_{\rm HK},
\end{equation}
where R$_{\rm phot}$ is the photospheric correction. It can be obtained from the color index $B-V$ using the function determined by \cite{Noyes1984}:
\begin{equation}
 \log R_{\rm phot}(B-V) = -4.898 + 1.918(B-V)^2 - 2.893 (B-V)^3. 
\end{equation}

We assume that the color index $B-V$ and thus the R$_{\rm phot}$ have not changed dramatically during the solar lifetime on the MS. According to \citet{Ribas_2010}, the effective temperature of the 1-Gyr old Sun was about 150 K lower than the present-day value. This would lead to a negligible change of the $B-V$ index.

According to \cite{Middelkoop1982}, the S-index can be calculated from the R$_{\rm HK}$ as:

\begin{equation}
    S =\frac{R_{\rm HK}} {1.34\times 10^{-4} \times C},
\end{equation}
where $C$ is given by:
\begin{equation}
 \log C (B-V) = 1.13(B-V)^3 - 3.91 (B-V)^2 + 2.84 (B-V) - 0.47.   
\end{equation}

Finally, using Eqs. (1), (2) and (4) we obtained the dependence of the S-index on the solar age t:

\begin{equation}
    S(t) =\frac{10^a + R_{\rm phot}} {1.34\times 10^{-4} \times C},
    \label{eq:S_T}
\end{equation}
where

\begin{equation}
   a=\frac{0.0534 - \log t} {1.92}.
\end{equation}

\begin{figure}
\resizebox{\hsize}{!}{\includegraphics [angle =90] {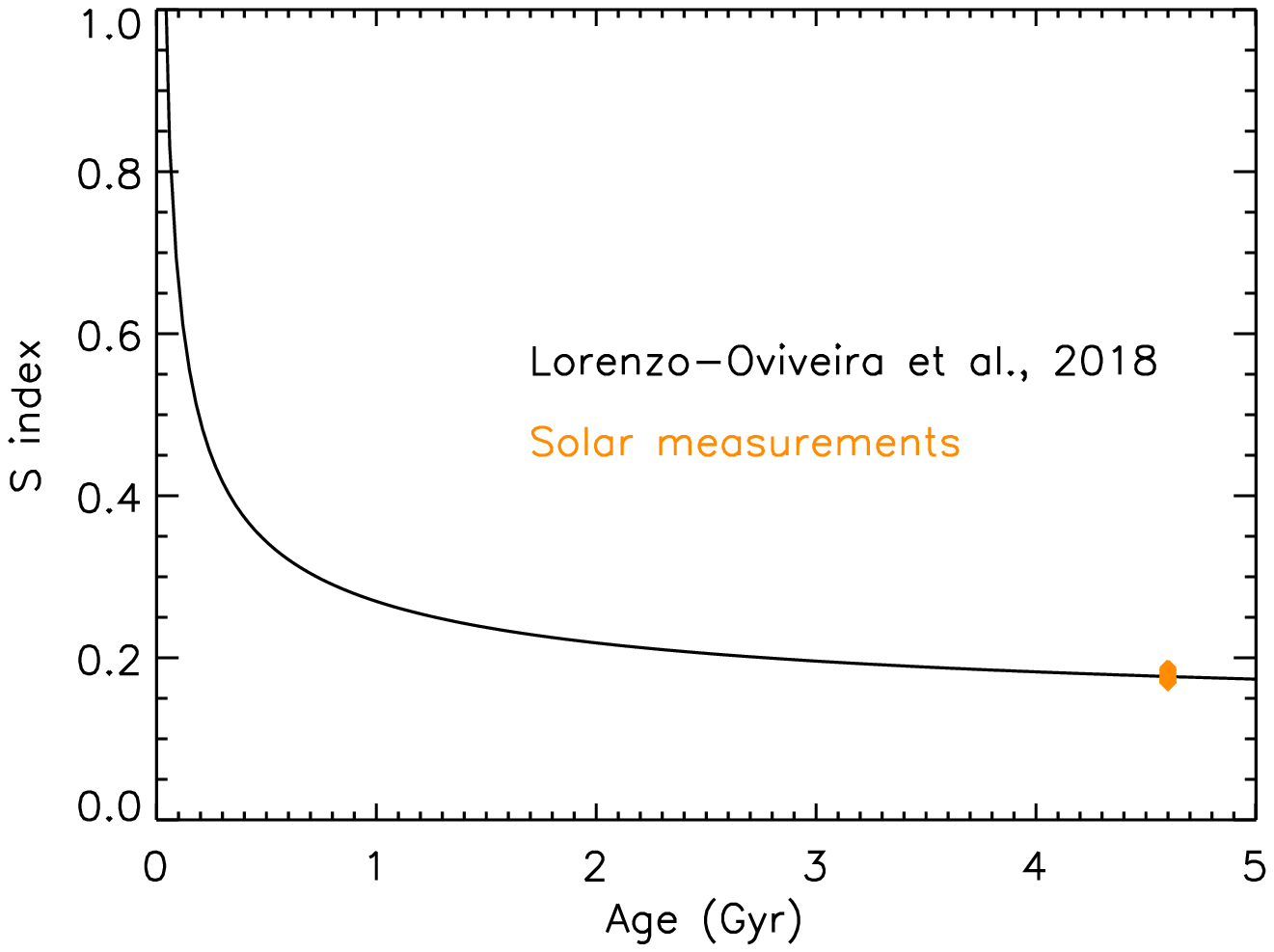}}
\caption{Evolution of the S-index according to \citet{LorenzoOliveira2018}. Orange diamonds represent annually averaged S-indexes values calculated with \ion{Ca} \relax II Sac Peak measurements.}
\label{fig:prof_act}
\end{figure}


 According to  Eq.~(\ref{eq:S_T}) the present-day Sun ($t=4.6$ Gyr) would have S-index equal to about 0.159. \citet{Shapiro2014} calculated annual mean values of the solar S-index for cycle 23 using the daily Sac Peak K-index K$_{\rm SP}$. They found that the value of the S-index averaged over cycle 23 was about 0.177. 
  Therefore, we scaled the S-index calculated with Eq.~(\ref{eq:S_T}) with the normalization coefficient of 1.11. The resulting dependence of the S-index on the solar age is plotted in Fig. 1.  We note that the exact value of the solar S-index depends on the employed calibration  \citep{Egeland2017} and the amplitude of the solar cycle also varies from cycle to cycle by more than 10\% \citep[see, e.g., Fig.~1 from][]{Shapiro2014} so that 11 \% difference between  two estimates is not surprising.  

\section{Model}
\subsection{Disk coverage by magnetic features}
Sunspots and faculae are the most prominent manifestations of the magnetic field in the solar photosphere \citep{Sami_B}. The sunspots are colder and darker than surrounding areas of the photosphere. They consist of an umbra, that is the central darkest part of the spot, and penumbra, that is a brighter part surrounding the umbra. Spots cover only a very small fraction of the present-day solar surface. For example, during the solar cycle 23 the coverage of the visible  solar disk by spots changed from about 0.3\%  in 2002, a period of high solar activity, to almost zero in 2008 when solar activity was low. Both numbers are annually averaged coverages. Faculae are usually brighter (especially in the cores of spectral lines and close to the solar limb) than the surrounding quiet Sun areas. In contrast to the sunspots, the coverage by faculae never goes to zero. During solar cycle 23 it changed from 0.27\% in 2002 to 0.02\% in 2008 \citep[see Fig.~1 in][]{Shapiro2014}.

 \citet{Shapiro2014} calculated the fractional coverage of the solar disk by sunspots $A_{\rm spot}$ and faculae $A_{\rm facular}$ as a function of the S-index:
\begin{equation}
  A_{\rm spots}(S) = 0.105 - 1.315\ S + 4.102\ S^2,   
\end{equation}
\begin{equation}
  A_{\rm faculae}(S) = -0.233 + 1.400\ S. 
\end{equation}
Eqs. (8,9) are obtained for the solar cycles 21-23, i.e. for the present-day level of solar activity. Nevertheless, calculations based on these functions were able to reproduce the activity cycle brightness variability of a variety of activity cycles measured in stars. Hence, we used Eqs. (6--9) to obtain the dependences of the solar disk area coverages by sunspots and faculae on the solar age (see Fig.~2). Due to the quadratic dependence of spot coverage on activity it decreases  with age faster than the facular coverage.

We note that \citet{LorenzoOliveira2018} observed stars at random phases of their activity cycles. Therefore the S-index given by Eq.~(\ref{eq:S_T}) and, consequently, the disk-area coverages obtained with such S-index (solid lines in Fig.~2) correspond to the cycle-averaged values. At the same time the crucial quantity for our calculations is the change of the disk-area coverages from minimum to maximum of the solar activity cycle (since such a change defines the amplitude of the cyclic irradiance variations). Following an approach of \cite{Egeland_PHD} we take the amplitude of the irradiance variations $\Delta S$ being proportional to the cycle-averaged value $S_{\rm mean}$. Consequently, we scale the amplitude of the cycle with the mean level of solar activity:
\begin{equation}
  \Delta S_{\rm var} (t)= \frac{\Delta S_{\rm var}^{p} \times S(t)}{ S_{\rm mean}^{\rm p} }, 
\end{equation}
where $S_{\rm mean}^{\rm p}$ is the mean level of the present-day S-index and $\Delta S_{\rm var}^{\rm p}$ is the amplitude of the present-day solar irradiance variations.  Both values have been calculated for the cycle 23 utilising S-index time series from \cite{Shapiro2014}. $S(t)$ is the dependence of the S-index on the solar age calculated with Eq. (6). Then the  S-index at the maximum $S_{\rm max}$ and minimum $S_{\rm min}$ of the solar activity cycle can be calculated as:
\begin{equation}
  S_{\rm max} (t)=\,S(t) + \Delta S_{\rm var}(t) /2
\end{equation}
\begin{equation}
  S_{\rm min} (t)=\,S(t) - \Delta S_{\rm var}(t) /2
\end{equation}

\begin{figure}
\resizebox{\hsize}{!}{\includegraphics [angle=90] {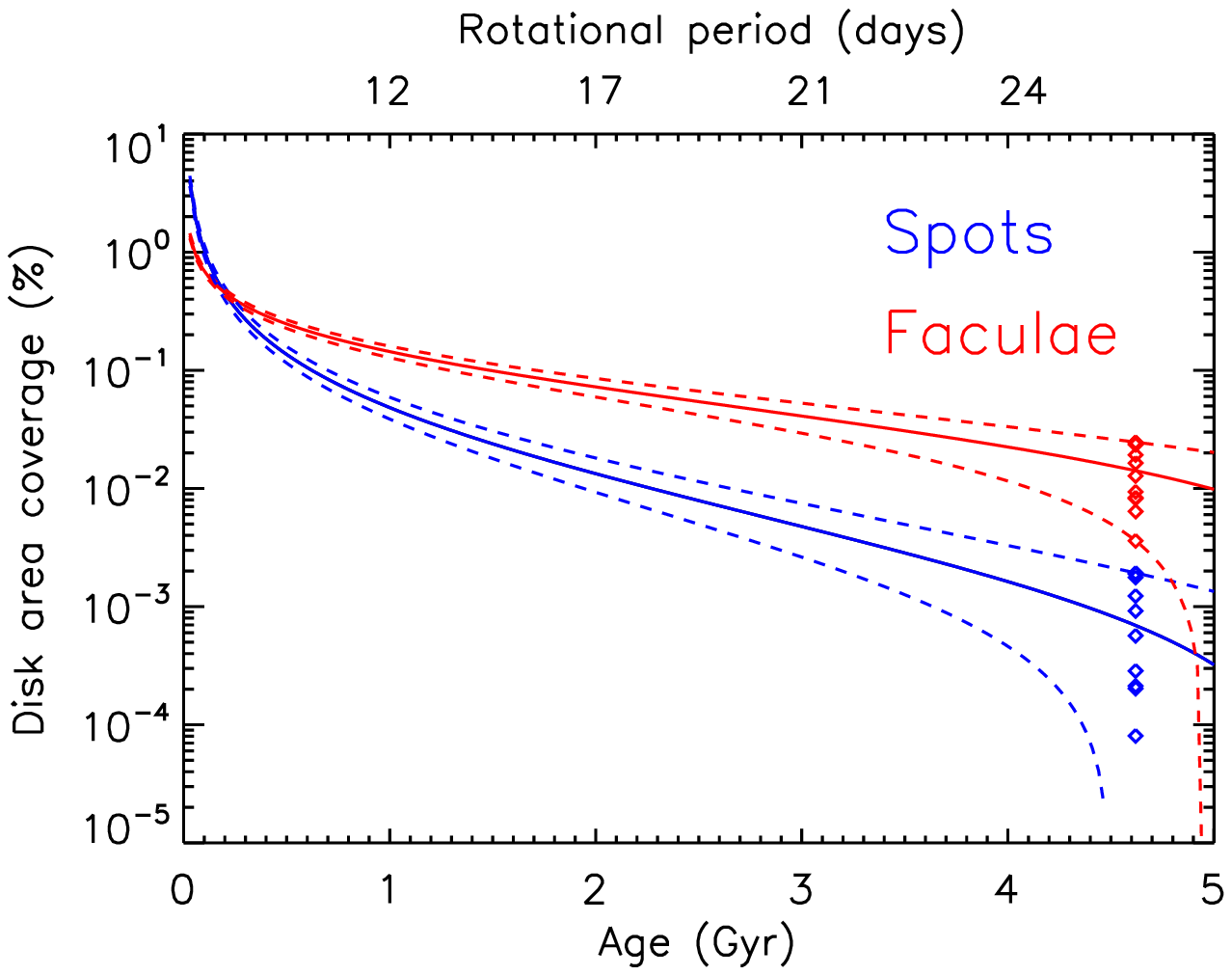}}
\caption{The dependences of solar disk-area coverages by sunspots (blue) and faculae (red) on solar age. Plotted are the dependences calculated with the S-index averaged over the activity cycle (solid lines) as well as the dependences obtained using the S-index at the minimum and maximum of the activity cycle (dashed lines below and above corresponding solid lines, respectively). The diamond symbols represent the disk-area coverages by sunspots (blue) and faculae (red) calculated with annual S-index values obtained from the Sac Peak \ion{Ca} \relax II measurements of solar cycle 23. The rotational period indicated along the top horizontal axis is calculated assuming that it is proportional to the square root of age \citep{Skumanich1972}.}
\label{fig:prof_act}
\end{figure}

The dependences calculated with Eqs.~(11--12) are indicated in Fig. 2 by dashed lines.

\subsection{Brightness variability}
In this section we use the solar disk area coverages by magnetic features obtained in Sect.~3.1. to calculate the variability of solar irradiance. For this we follow the SATIRE approach \citep[here SATIRE stands for Spectral And Total Irradiance Reconstruction, see][]{Fligge2000, Krivova2003, Krivova2011} generalized by \cite{Shapiro2014} to calculate variability of Sun-like stars. Since the goal of this study is to reconstruct the activity cycle in SSI we extend the \cite{Shapiro2014} approach to compute SSI variability as a function of wavelength in the 160--2430 nm spectral range.

SATIRE attributes the SSI variability to the time-dependent contributions from bright faculae and dark sunspot. In this study we limit ourselves to modelling the variability on the timescale of solar activity cycle and do not consider variations brought about by transits of specific magnetic features. Therefore we follow the approach by \cite{Knaack2001} and \cite{Shapiro2014} and assume that spots are  distributed in the activity belts between $\pm$5$^\circ$ and $\pm$30$^\circ$ latitudes while faculae are distributed between $\pm$5$^\circ$ and $\pm$40$^\circ$ latitudes. We neglect the change of the spatial distribution of the solar magnetic features with the solar age. This is a reasonable assumption for activity levels up to 4 times of the present Sun \citep{Emre2018}, i.e. back to solar age of about 0.5 Gyr. In this case the active component of the SSI depends only on the total area coverage by faculae and spots (since the area distribution on the disk is fixed) which is, in turn, given by the S-index. Consequently, SSI can be written as a function of the wavelength and S-index:
\begin{equation}
F(\lambda, S) = F_{\rm quiet}(\lambda) + \Delta F_{\rm u}(\lambda,S) + \Delta F_{\rm p}(\lambda,S) + \Delta F_{\rm f}(\lambda,S).
\end{equation}
Here $F_{\rm quiet}(\lambda)$ is the irradiance from the full disk quiet Sun, i.e. irradiance from the Sun without any magnetic features. $\Delta F_{\rm u}(\lambda,S)$, $\Delta F_{\rm p}(\lambda,S)$, and $\Delta F_{\rm f}(\lambda,S)$ are changes in the SSI from spot umbrae, spot penumbrae, and faculae, respectively. They can be written as:
\begin{equation}
 \Delta F_i\,(\lambda, S) = \iint\limits_{\rm solar \, disk} a_i (S,\vec{n} ) \left[ I_i(\lambda,\vec{n}) - I_{\rm quiet}(\lambda,\vec{n}) \right] \rm d\Omega ,
\end{equation}
where $I_{\rm quiet}(\lambda,\vec{n})$ is the intensity from the quiet Sun at wavelength $\lambda$ in the direction of the line of sight $\vec{n}$, $I_{i}(\lambda,\vec{n})$ is the intensity from magnetic features, with  the index $i$ refers to either u, p, or f, and   $\rm d\Omega$ is the element of solid angle around the direction $\vec{n}$. The quantity $a_i(S,\vec{n})$ is the fractional coverage of the solar disk by the active component $i$ in the direction $\vec{n}$; it takes values between zero and unity within the corresponding activity belt (i.e. the region containing spots and faculae) and is zero outside. The intensities of the quiet Sun, faculae, umbrae, and penumbrae have been calculated by \cite{Unruh1999} with the ATLAS9 code \citep{kurucz1992,ATLAS9_CK} as functions of the wavelengths and of the cosine of the angle between the direction to the observer and the local stellar radius.

We define the magnetic component of solar brightness as:
\begin{equation}
  R(\lambda,S) = \frac{\Delta F_{\rm u}(\lambda,S) + \Delta F_{\rm p}(\lambda,S) + \Delta F_{\rm f}(\lambda,S)}{F_{\rm quiet}(\lambda)}.  
 \end{equation}
 
We also consider the spot and facular components separately:
\begin{equation}
  R_{\rm f}(\lambda,S) = \frac{\Delta F_{\rm f}(\lambda,S)}{F_{\rm quiet}(\lambda)}, 
\end{equation}
\begin{equation}
  R_{\rm s}(\lambda,S) = \frac{\Delta F_{\rm u}(\lambda,S) + \Delta F_{\rm p}(\lambda,S)}{F_{\rm quiet}(\lambda)}.  
\end{equation}

 Because the solar disk coverages by spots and faculae do not drop to zero during activity minima (see Fig. 2) the contribution of magnetic features to solar irradiance derived from Eqs. (13--17) also never reach zero. The change of the magnetic component from the solar cycle minimum to maximum defines the  amplitude of the cyclic SSI variability:
\begin{equation}
  D(\lambda,S) = R(\lambda,S_{\rm max}) - R(\lambda,S_{\rm min}).
\end{equation}
Here S is the mean value of the S-index over the solar cycle, while $S_{\rm max}$ and $S_{\rm min}$ are the S-index values during the activity maximum and minimum, respectively. Combining Eq. (18) with Eqs.~(6)~and~(10--12) we can reconstruct the amplitude of the cyclic SSI variability as a function of solar age.

 We note that to get the $R(\lambda,S)$ values, the absolute changes of irradiance due to magnetic features (nominators in Eqs.~15--17)  are divided by the irradiance from the quiet Sun, $F_{\rm quiet}$. Hence, the amplitude of the cyclic SSI variability, $ D(\lambda,S) $ given by Eq.~(18) is dimensionless. For simplicity we will refer to  $ D(\lambda,S) $ as to the relative variability (even though the relative variability would be, strictly speaking, obtained by dividing the absolute change of irradiance by the mean value of irradiance during the activity cycle.).

 In the present study we do not aim to model the exact time profile of the SSI change over the activity cycle, but restrict ourselves to modelling its amplitude. In addition, by taking the amplitude of the solar cycle 23 in the S-index  as a reference for the present-day solar variability (see  description of Eq.~10 in Sect.~3.1), we are considering cycles of average strength (in the last four centuries both, much stronger and much weaker solar activity cycles have been observed).

\section{Results}
 In this section we show how the amplitude of solar irradiance variability on the activity cycle timescale depends on the solar age. Sect.~4.1~and~4.2 present calculations of the variability in TSI and several spectral passbands, respectively, while Sect.~4.3 focuses on the variability at high spectral resolution.

\subsection{Variability of TSI}
We first calculate the amplitude of the cyclic variability in the 160--2430 nm spectral domain (hereafter referred to as the TSI variability) as well as its facular and spot components. This is done using Eqs.~(15--17). 

Figure~3 presents the dependence of the relative amplitude of the TSI variability (i.e. the difference between TSI values at cycle maximum and minimum normalised to the TSI value from the quiet Sun, see Sect.~3.2) on time.  One can see that the TSI variability is out-of-phase  with the activity for the Sun at ages less than about 2.8 Gyr and becomes in-phase afterwards. The out-of-phase variability of the young Sun implies that TSI in the minimum of the activity cycle is larger than that in the maximum, i.e. the variability is spot-dominated. Likewise, the in-phase variability when the Sun is older than 2.8 Gyr means that TSI during activity minima is smaller than that during the maxima and the variability is faculae-dominated. The change of the variability from spot- to faculae-dominated for the 2.8-Gyr Sun is in line with the observations of \cite{Lockwood_2007,Radick2018,Timo2019}, who found that a similar transition between spot- and faculae-dominated regimes happens for stars more active than the Sun. 

The TSI variability of the present Sun is faculae-dominated. Going back in time the solar S-index increases (see Fig.~1). Consequently, the solar disk coverages by both, spots and faculae, also increase. However, such an increase is stronger for spots than for faculae  (see Eqs. 8-9 and Fig.~2). Consequently, the spot component of the variability increases faster than the facular component (the red line in Fig. 3 is almost horizontal at the solar age, while the blue curve is significantly inclined). This leads to a stronger compensation between facular and spot components of the variability and to the drop of the total variability. All in all, the  amplitude of the variability decreases despite the increase of the solar magnetic activity. We note that such a compensation effect mainly happens on the timescale of the activity cycle \citep[][]{Shapiro2016}. We expect that the TSI variability on the rotational timescale, which is spot-dominated for the present Sun, monotonically increases with decreasing solar age. The calculation of the rotational variability demands the knowledge of the sizes and spatial distribution of individual magnetic features and thus is outside the scope of the present study. 

The  amplitude of the variability returned by Eq.~(16) reaches zero at 2.8 Gyr, i.e. the TSI values at the maximum and minimum of the activity cycle becomes equal. This implies that in-phase UV variability is compensated by the out-of-phase variability in the visible and infrared spectral domains. We note that zero amplitude of the TSI variability does not mean that the TSI will be constant over the whole activity cycle. The ratio between solar disk coverages by spots and faculae changes over the cycle \citep[see, e.g.][]{Shapiro2014} so that spots and faculae will not totally cancel each other at intermediate cycle phases.

The variability of the Sun younger than 2.8 Gyr rapidly increases with decreasing of the solar age. The dependence on age is much stronger for the spot-dominated Sun than for the facular-dominated Sun (compare ``faculae-dominated'' and ``spot-dominated'' parts of the black curve in Fig.~3). This is because the spot component of the variability increases with activity faster than the facular component. Interestingly, the amplitude of the variability of the 600-Myr Sun reaches 1\%, i.e. it is one order of an magnitude larger than the variability of the present Sun.


\begin{figure}
\resizebox{\hsize}{!}{\includegraphics [angle=90] {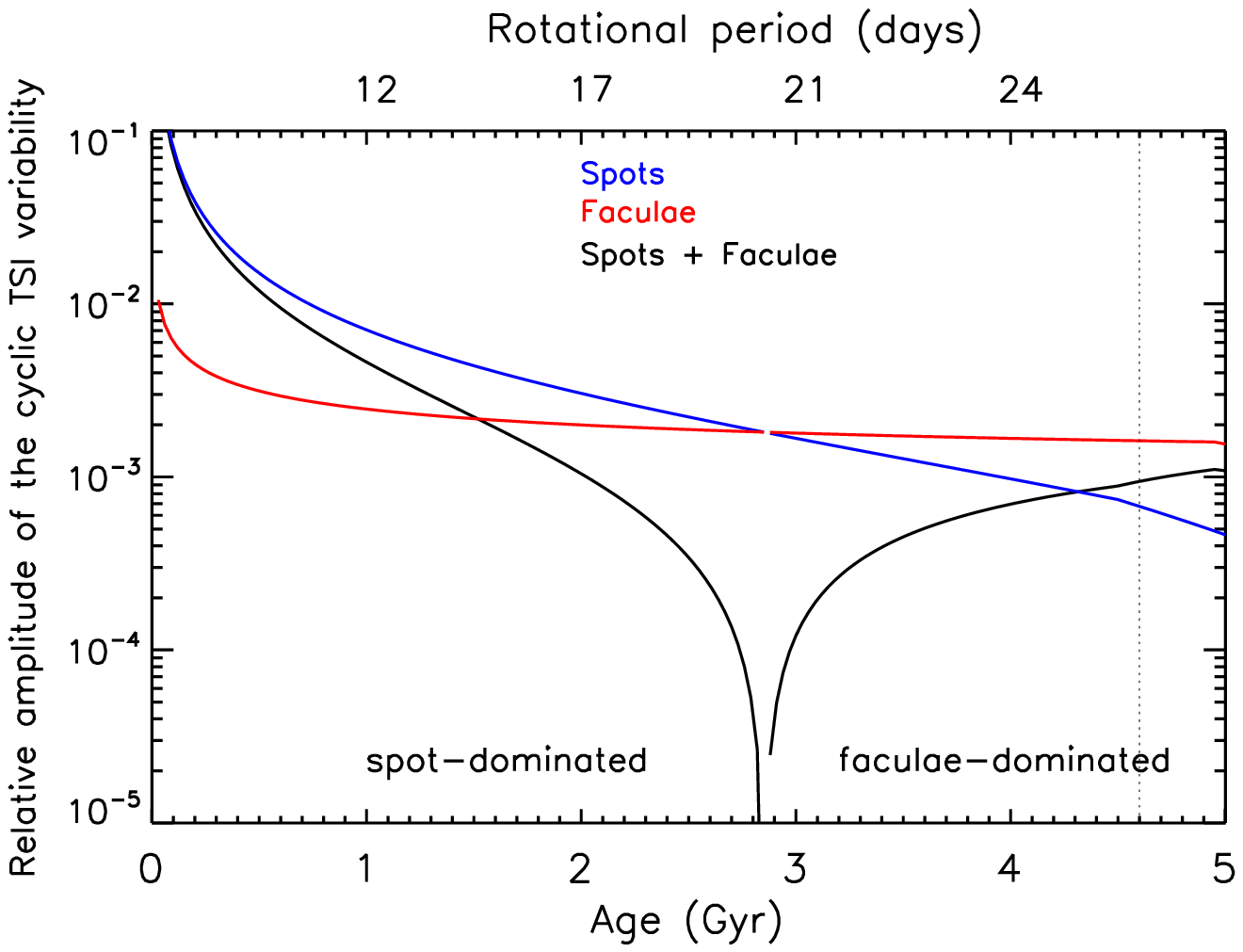}}
\caption{The dependence of the solar cycle TSI variability  on the solar age. The value is calculated using Eq.(6, 18) and equal to $D(\lambda,S)$ integrated over spectral range 160-2430 nm. We show the total amplitude (black), its facular (red) and spot (blue) components in absolute values. The spot contribution is negative. The facular contribution is positive. Thus the total amplitude can be understood as a difference between its facular and spot components (in absolute values). The dashed vertical black line points to the present solar age. The TSI variability back to about 2.8 Gyr is faculae-dominated, while the variability of the younger Sun is spot-dominated. The corresponding parts of the black curve are designated "faculae-dominated" and "spot-dominated", respectively. The period indicated along the top X-axis is calculated assuming that it is proportional to the square root of the age. }
\label{fig:prof_act}
\end{figure}

\subsection{ Variability in spectral bands}

Figure 4 illustrates the behaviour of solar irradiance variability in five spectral bands that are important for the present-day photochemistry in the Earth's atmosphere \citep{Ermolli2013}: 160--210 nm, 210--400 nm, 400--700 nm, 700--1000 nm, and 1000--2430 nm. We note that the ATLAS9 code used for calculating spectra of the quiet Sun and magnetic features (see Sect.~3.2) operates under the assumption of the Local Thermodynamic Equilibrium which fails in the UV spectral range. Consequently, the amplitude of the solar irradiance variability below 180 nm was corrected by \cite{yeoetal2014} using the irradiance measurements by SORCE/SOLSTICE instrument. The absence of such a correction in our calculations might affect the results for the 160--210 nm band \citep[see][for a detailed discussion of the effects caused by the deviations from the LTE]{Rinat2019}. 


\begin{figure}
\resizebox{\hsize}{!}{\includegraphics {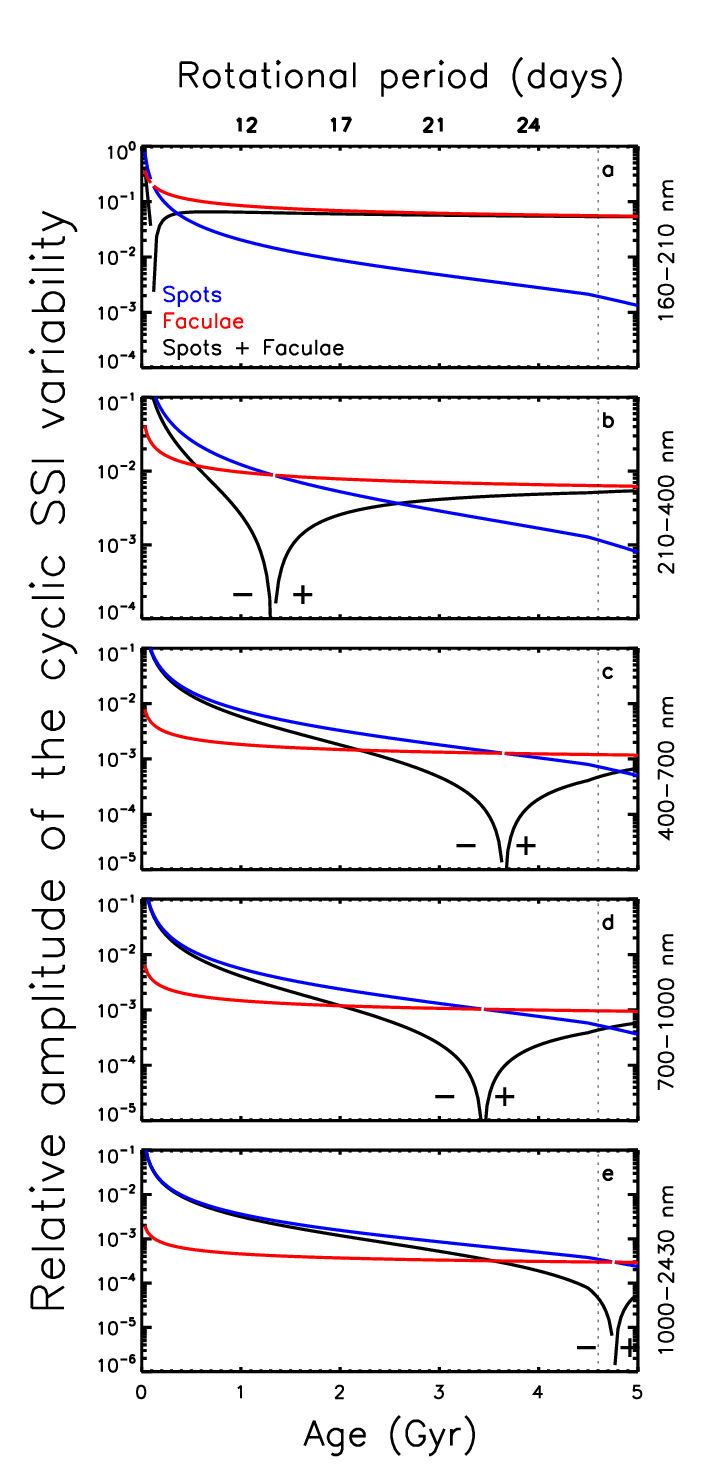}}
\caption{The same as Figure 3 but for the variability in 160--210 nm (a), 210--400 nm (b), 400--700 nm (c), 700--1000 nm (d), and 1000--2430 nm (e) spectral domains. The spot- and faculae-dominated parts of the black curve are designed by "$-$" and "$+$" sign, respectively.}
\label{fig:prof_act}
\end{figure}

Similarly to the TSI behaviour, the variability in spectral bands shows non-monotonous dependence on the solar age. The activity level and solar age corresponding to the transition from spots to faculae domination depends on the relative contrasts of spots and faculae and, hence, on the wavelength. Faculae are especially bright in the UV. Therefore, the facular component gets stronger in the 210--400 nm and 160--210 nm bands and the transition from  spot- to faculae-dominated regimes happens at higher activity, i.e. for a younger Sun (compare different panels in Fig.~4).

 Interestingly, the amplitude of the 160--210 nm flux variability shows a rather weak dependence on age down to the solar age of about 600 Myr (see Fig.~4a). This is because variability is mainly brought about by faculae after 2 Gyr and thus it scales almost linearly with the S-index, which only increases by 21\% down to 2 Gyr relative to the present-day value. Interestingly, the spot component which starts to contribute roughly at 2 Gyr   further slows the growth of the variability. For example, the 1-Gyr old Sun was only  30\% more variable in this domain than the present-day Sun. 
We note that the change of the variability in TSI and other spectral bands shown in Fig.~4 over the last 4 Gyr has been significantly higher than that in the 160--210 nm band. This is due to the compensation effect between spot and facular contributions to the variability. In other words, even a small change of facular and spot contributions in these passbands lead to a substantial change of the difference between them (see Fig.~3~and~4). 
For example, the variability in  210-400 nm spectral band at 1.5 Gyr was about 5 times larger than for the present Sun (to be compared with 14\% increase in 160--210 nm band).

\begin{figure*}
\resizebox{\hsize}{!}{\includegraphics [angle=90] {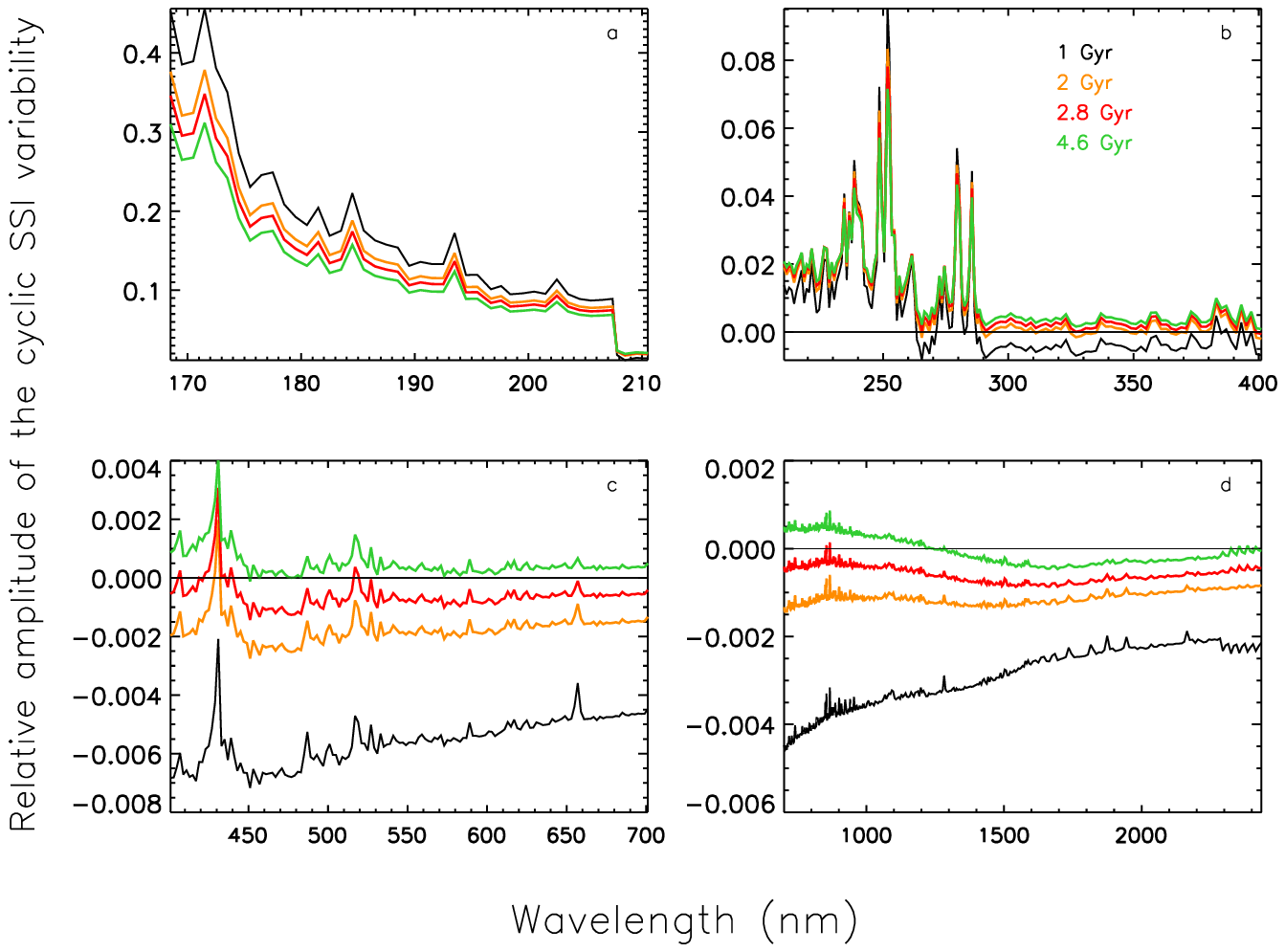}}
\caption{The amplitude of the solar cycle in irradiance variations as a function of wavelength in the 168--210 nm (panel a), 210--400 nm (panel b), 400--700 nm (panel c), and 700--2430 nm (panel d) spectral bands. Shown are the amplitudes for the solar age of 1 Gyr (black), 2 Gyr (orange), 2.8 Gyr (red) and 4.6 Gyr (green). The horizontal black line shows the Sun at the zero level of the variability.}
\label{fig:prof_act}
\end{figure*}

\subsection{Variability at high spectral resolution}
In Sect.~4.1 we looked at the amplitude of the variability in several spectral bands as a function of solar age. In this section we present the amplitude of the variability as a function of wavelength for selected solar ages. Figure~5 shows calculations for the 1-Gyr, 2-Gyr, 2.8-Gyr, and 4.6-Gyr (i.e. present) Sun. The spectral dependences have a rather rich profile due to the effect of Fraunhofer lines \citep[see][]{Unruh1999, Shapiroetal2015}.

 One can see that the amplitude of the cyclic variability does not show a strong dependence on the solar age in the 168--210 nm spectral domain (see Fig.~5a). 
This is because the variability in this spectral domain is dominated by the faculae component and scales linearly with the solar S-index (see discussion in Sect.~3.2 and Fig.~4a). Interestingly, the behaviour of the variability in the pronounced peaks in Fig.~5b is similar to that of the variability in the 
168--210 nm spectral domain. The increase of the variability indicated by these peaks is caused by strong spectral lines which amplify the facular contrast \citep{Shapiroetal2015}. Consequently, the variability in the wavelengths corresponding to these peaks is dominated by faculae and its amplitude monotonously increases with the decrease of the solar age (note the opposite behaviour from the variability outside of the spectral range of the peaks). We note, however, that \cite{Unruh1999} spectra utilised in our calculations have been computed ignoring chromospheric contribution and non-LTE effects. Both these shortcomings are expected to affect the amplitude of the variability in strong spectral lines \cite[see, e.g., discussion in][]{Rinat2019} and addressing them is outside of the scope of the present study. 

The spot component of the variability starts to play an important role longwards Mg II doublet at 280 nm. In particular, variability of the 1 Gyr Sun is spot-dominated there, i.e. the amplitude of the relative variability is negative (see Figs.~5b--d). As the Sun ages both spot and facular components of the variability decrease but the spot component decreases faster. As a result spot and facular components almost cancel each other in between 280 nm and 380 nm for the 2 Gyr Sun. The variability in this spectral domain becomes faculae-dominated for the 2.8 Gyr and 4.6 Gyr Sun. Interestingly, despite the amplitude of both facular and spot components decrease with solar age the total variability (given by the difference between facular and spot components) increases. This is the same effect which causes the increase of the variability after the transition from spot- to faculae-dominated regimes discussed in Sect.~4.1~and~4.2 (see also Figs.~3~and~4). 

Figs.~5cd show that while the variability of the present Sun is faculae-dominated up to about 1200 nm, the variability of the 2.8 Gyr Sun becomes spot-dominated longwards the strong peak at 430 nm caused by the CH G-band.

\section{Summary}

We have employed the SATIRE approach to investigate how the amplitude of solar irradiance variability on the timescale of the activity cycle might have changed with the solar age. 
We have separately modeled the contributions from spots and faculae and found that both of them monotonically decrease with the age. 

The key ingredient of our model is the observed dependence of the ratio between spot and facular coverages on solar activity level. It causes the  difference in pattern of the spot and facular increase with solar age which results in a non-monotonous dependence of the solar irradiance variability on the age. The amplitude of the cyclic TSI variability decreased till the solar age of 2.8 Gyr. The amplitude was weakest at around 2.8 Gyr. Then it started to increase but the amplitudes of the present-day Sun is still about 10 times smaller than the value for the 600 Myr old Sun.

 During the periof of the weakest amplitude at 2.8 Gyr the faculae-dominated change of the UV irradiance was almost fully compensated by the spot-dominated change of the irradiance in visible and infrared spectral domains. The faculae-dominated change is in-phase with solar activity while the spot-dominated change is out-of-phase.

We calculated the variability in TSI and in several spectral domains: 160--210, 210--400, 400--700, 700--1000 and 1000--2430 nm. Our computations show that the age at which transition between the spot and facular domination happens depends on the wavelength. This is because the contrast of magnetic features depends on the wavelength. We have also obtained the wavelength dependences of the variability for solar ages of 1, 2, 3.8, and 4.6 Gyr.


 A stronger cyclic variability of solar brightness could lead to a stronger response of planetary atmospheres. Hence, our calculations might be of interest for  the paleoclimate research and for studying the historical conditions of the atmospheres of other planets in the solar system. The interest to such studies is being raised by the advent of the new data, e.g.  the main goal of the MAVEN mission \citep{Jakosky_2015} launched in 2013 is to understand the Martian environment in the past.

\begin{acknowledgements}
We would like to thank Frederico Spada and Nadiia Kostogryz for useful discussions. The research performed for this paper was partially supported by the German space agency (Deutsches Zentrum für Luft- und Raumfahrt) under PLATO Data Center grant 50OO1501. A.I.S. acknowledges  funding  from  the European Research Council under the European Union Horizon 2020 research and innovation programme (grant agreement No. 715947). L.G. acknowledges partial support from ERC Synergy grant WHOLE SUN 810218.
\end{acknowledgements}

\bibliographystyle{aa}

\begin{thebibliography}{48}
\expandafter\ifx\csname natexlab\endcsname\relax\def\natexlab#1{#1}\fi

\bibitem[{{Baliunas} {et~al.}(1995){Baliunas}, {Donahue}, {Soon}, {Horne},
  {Frazer}, {Woodard-Eklund}, {Bradford}, {Rao}, {Wilson}, {Zhang}, {Bennett},
  {Briggs}, {Carroll}, {Duncan}, {Figueroa}, {Lanning}, {Misch}, {Mueller},
  {Noyes}, {Poppe}, {Porter}, {Robinson}, {Russell}, {Shelton}, {Soyumer},
  {Vaughan}, \& {Whitney}}]{Baliunas1995}
{Baliunas}, S.~L., {Donahue}, R.~A., {Soon}, W.~H., {et~al.} 1995, \apj, 438,
  269

\bibitem[{Balogh {et~al.}(2014)Balogh, Hudson, Petrovay, \& von
  Steiger}]{Balogh2014}
Balogh, A., Hudson, H.~S., Petrovay, K., \& von Steiger, R. 2014, {\spsr}, 186,
  1

\bibitem[{{Brun} \& {Browning}(2017)}]{Brun_LR}
{Brun}, A.~S. \& {Browning}, M.~K. 2017, \lrsph, 14, 4

\bibitem[{{Castelli} \& {Kurucz}(1994)}]{ATLAS9_CK}
{Castelli}, F. \& {Kurucz}, R.~L. 1994, \aap, 281, 817

\bibitem[{{Charbonneau}(2010)}]{Charbonneau2010}
{Charbonneau}, P. 2010, {\lrsph}, 7, 91

\bibitem[{{Egeland}(2017)}]{Egeland_PHD}
{Egeland}, R. 2017, PhD thesis, Montana State University, Bozeman, Montana, USA

\bibitem[{{Egeland} {et~al.}(2017){Egeland}, {Soon}, {Baliunas}, {Hall},
  {Pevtsov}, \& {Bertello}}]{Egeland2017}
{Egeland}, R., {Soon}, W., {Baliunas}, S., {et~al.} 2017, \apj, 835, 25

\bibitem[{{Ermolli} {et~al.}(2013){Ermolli}, {Matthes}, {Dudok de Wit},
  {Krivova}, {Tourpali}, {Weber}, {Unruh}, {Gray}, {Langematz}, {Pilewskie},
  {Rozanov}, {Schmutz}, {Shapiro}, {Solanki}, \& {Woods}}]{Ermolli2013}
{Ermolli}, I., {Matthes}, K., {Dudok de Wit}, T., {et~al.} 2013, {\atcp}, 13,
  3945

\bibitem[{{Fligge} {et~al.}(2000){Fligge}, {Solanki}, \& {Unruh}}]{Fligge2000}
{Fligge}, M., {Solanki}, S.~K., \& {Unruh}, Y.~C. 2000, {\aap}, 353, 380

\bibitem[{Floyd {et~al.}(2003)Floyd, Rottman, Deland, \& Pap}]{Floyd2003}
Floyd, L., Rottman, G., Deland, M., \& Pap, J. 2003, {Solar variability as an
  input to the Earth's environment}, 535, 195

\bibitem[{Galarza {et~al.}(2019)Galarza, MelŽndez, Lorenzo-Oliveira, Valio,
  Reggiani, Carlos, Ponte, Spina, Haywood, \& Gandolfi}]{Galanza2019}
Galarza, J.~Y., Melendez, J., Lorenzo-Oliveira, D., {et~al.} 2019, Monthly
  Notices of the Royal Astronomical Society: Letters

\bibitem[{{Guinan} {et~al.}(2003){Guinan}, {Ribas}, \& {Harper}}]{Guinan2003}
{Guinan}, E.~F., {Ribas}, I., \& {Harper}, G.~M. 2003, {\apj}, 594, 561

\bibitem[{{Guedel}(2007)}]{Guebel2007}
{Guedel}, M. 2007, {\lrsph}, 4, 137

\bibitem[{{Hall} {et~al.}(2009){Hall}, {Henry}, {Lockwood}, {Skiff}, \&
  {Saar}}]{halletal2009}
{Hall}, J.~C., {Henry}, G.~W., {Lockwood}, G.~W., {Skiff}, B.~A., \& {Saar},
  S.~H. 2009, \aj, 138, 312

\bibitem[{{Hood} {et~al.}(2015){Hood}, {Misios}, {Mitchell}, {Rozanov}, {Gray},
  {Tourpali}, {Matthes}, {Schmidt}, {Chiodo}, \& {Thi{\'e}blemont}}]{Hood2015}
{Hood}, L.~L., {Misios}, S., {Mitchell}, D.~M., {et~al.} 2015, Quarterly
  Journal of the Royal Meteorological Society, 141, 2670

\bibitem[{{I{\c s}{\i}k} {et~al.}(2018){I{\c s}{\i}k}, {Solanki}, {Krivova}, \&
  {Shapiro}}]{Emre2018}
{I{\c s}{\i}k}, E., {Solanki}, S.~K., {Krivova}, N.~A., \& {Shapiro}, A.~I.
  2018, \aap, 620, A177

\bibitem[{{Jakosky} {et~al.}(2015){Jakosky}, {Lin}, {Grebowsky}, {Luhmann},
  {Mitchell}, {Beutelschies}, {Priser}, {Acuna}, {Andersson}, {Baird}, {Baker},
  {Bartlett}, {Benna}, {Bougher}, {Brain}, {Carson}, {Cauffman}, {Chamberlin},
  {Chaufray}, {Cheatom}, {Clarke}, {Connerney}, {Cravens}, {Curtis}, {Delory},
  {Demcak}, {DeWolfe}, {Eparvier}, {Ergun}, {Eriksson}, {Espley}, {Fang},
  {Folta}, {Fox}, {Gomez-Rosa}, {Habenicht}, {Halekas}, {Holsclaw}, {Houghton},
  {Howard}, {Jarosz}, {Jedrich}, {Johnson}, {Kasprzak}, {Kelley}, {King},
  {Lankton}, {Larson}, {Leblanc}, {Lefevre}, {Lillis}, {Mahaffy}, {Mazelle},
  {McClintock}, {McFadden}, {Mitchell}, {Montmessin}, {Morrissey}, {Peterson},
  {Possel}, {Sauvaud}, {Schneider}, {Sidney}, {Sparacino}, {Stewart}, {Tolson},
  {Toublanc}, {Waters}, {Woods}, {Yelle}, \& {Zurek}}]{Jakosky_2015}
{Jakosky}, B.~M., {Lin}, R.~P., {Grebowsky}, J.~M., {et~al.} 2015, \ssr, 195, 3

\bibitem[{{Knaack} {et~al.}(2001){Knaack}, {Fligge}, {Solanki}, \&
  {Unruh}}]{Knaack2001}
{Knaack}, R., {Fligge}, M., {Solanki}, S.~K., \& {Unruh}, Y.~C. 2001, {\aap},
  376, 1080

\bibitem[{{Krivova} {et~al.}(2003){Krivova}, {Solanki}, {Fligge}, \&
  {Unruh}}]{Krivova2003}
{Krivova}, N.~A., {Solanki}, S.~K., {Fligge}, M., \& {Unruh}, Y.~C. 2003,
  {\aap}, 399, L1

\bibitem[{{Krivova} {et~al.}(2010){Krivova}, {Vieira}, \&
  {Solanki}}]{Krivova2011}
{Krivova}, N.~A., {Vieira}, L. E.~A., \& {Solanki}, S.~K. 2010, {\jgr}, 115,
  A12112

\bibitem[{{Kurucz}(1992)}]{kurucz1992}
{Kurucz}, R.~L. 1992, Rev. Mex. de Astron. Astrof., 23, 181

\bibitem[{{Lachaume} {et~al.}(1999){Lachaume}, {Dominik}, {Lanz}, \&
  {Habing}}]{Lachaume1999}
{Lachaume}, R., {Dominik}, C., {Lanz}, T., \& {Habing}, H.~J. 1999, {\aap},
  348, 897

\bibitem[{{Lockwood} {et~al.}(2007){Lockwood}, {Skiff}, {Henry}, {Henry},
  {Radick}, {Baliunas}, {Donahue}, \& {Soon}}]{Lockwood_2007}
{Lockwood}, G.~W., {Skiff}, B.~A., {Henry}, G.~W., {et~al.} 2007, \apjs, 171,
  260

\bibitem[{{Lorenzo-Oliveira} {et~al.}(2018){Lorenzo-Oliveira}, {Freitas},
  {MelŽndez}, {Bedell}, {Ram'rez}, {Bean}, {Asplund}, {Spina}, {Dreizler},
  {Alves-Brito}, \& {Casagrande}}]{LorenzoOliveira2018}
{Lorenzo-Oliveira}, D., {Freitas}, F.~C., {Melendez}, J., {et~al.} 2018,
  {\aap}, 619, 10

\bibitem[{{Mamamjek} \& {Hillenbrand}(2008)}]{Mamajek2008}
{Mamamjek}, E.~E. \& {Hillenbrand}, L.~A. 2008, {The Astronomical Journal},
  687, 1264

\bibitem[{{Maycock} {et~al.}(2018){Maycock}, {Matthes}, {Tegtmeier}, {Schmidt},
  {Thi{\'e}blemont}, {Hood}, {Akiyoshi}, {Bekki}, {Deushi}, {J{\"o}ckel},
  {Kirner}, {Kunze}, {Marchand}, {Marsh}, {Michou}, {Plummer}, {Revell},
  {Rozanov}, {Stenke}, {Yamashita}, \& {Yoshida}}]{Maycock2018}
{Maycock}, A.~C., {Matthes}, K., {Tegtmeier}, S., {et~al.} 2018, \atcp, 18,
  11323

\bibitem[{{Meehl} {et~al.}(2009){Meehl}, {Arblaster}, {Matthes}, {Sassi}, \&
  {van Loon}}]{Meehl2009}
{Meehl}, G.~A., {Arblaster}, J.~M., {Matthes}, K., {Sassi}, F., \& {van Loon},
  H. 2009, Science, 325, 1114

\bibitem[{{Metcalfe} {et~al.}(2016){Metcalfe}, {Egeland}, \& {van
  Saders}}]{Metcalfe2016}
{Metcalfe}, T.~S., {Egeland}, R., \& {van Saders}, J. 2016, \apj, 826, L2

\bibitem[{{Middelkoop}(1982)}]{Middelkoop1982}
{Middelkoop}, F. 1982, {\aap}, 107, 31

\bibitem[{{Noyes} {et~al.}(1984){Noyes}, {Hartmann}, {Baliunas}, {Duncan}, \&
  {Vaughan}}]{Noyes1984}
{Noyes}, R.~W., {Hartmann}, L.~W., {Baliunas}, S.~L., {Duncan}, D.~K., \&
  {Vaughan}, A.~H. 1984, {\apj}, 279, 763

\bibitem[{{Parker}(1955)}]{Parker1955}
{Parker}, E.~N. 1955, \apj, 122, 293

\bibitem[{{Radick} {et~al.}(2018){Radick}, {Lockwood}, {Henry}, {Hall}, \&
  {Pevtsov}}]{Radick2018}
{Radick}, R.~R., {Lockwood}, G.~W., {Henry}, G.~W., {Hall}, J.~C., \&
  {Pevtsov}, A.~A. 2018, \apj, 855, 75

\bibitem[{{Radick} {et~al.}(1998){Radick}, {Lockwood}, {Skiff}, \&
  {Baliunas}}]{radicketal1998}
{Radick}, R.~R., {Lockwood}, G.~W., {Skiff}, B.~A., \& {Baliunas}, S.~L. 1998,
  \apjs, 118, 239

\bibitem[{{Reiners} \& {Mohanty}(2012)}]{Reiners2012}
{Reiners}, A. \& {Mohanty}, S. 2012, \apj, 746, 43

\bibitem[{{Reinhold} {et~al.}(2019){Reinhold}, {Bell}, {Kuszlewicz}, {Hekker},
  \& {Shapiro}}]{Timo2019}
{Reinhold}, T., {Bell}, K.~J., {Kuszlewicz}, J., {Hekker}, S., \& {Shapiro},
  A.~I. 2019, \aap, 621, A21

\bibitem[{{Ribas}(2010)}]{Ribas_2010}
{Ribas}, I. 2010, in IAU Symposium, Vol. 264, Solar and Stellar Variability:
  Impact on Earth and Planets, ed. A.~G. {Kosovichev}, A.~H. {Andrei}, \& J.-P.
  {Rozelot}, 3--18

\bibitem[{{Shapiro} {et~al.}(2014){Shapiro}, {Solanki}, {Krivova}, {Schmutz},
  {Ball}, {Knaack}, {Rozanov}, \& {Unruh}}]{Shapiro2014}
{Shapiro}, A.~I., {Solanki}, S.~K., {Krivova}, N.~A., {et~al.} 2014, {\aap},
  569, 18

\bibitem[{{Shapiro} {et~al.}(2015){Shapiro}, {Solanki}, {Krivova}, {Tagirov},
  \& {Schmutz}}]{Shapiroetal2015}
{Shapiro}, A.~I., {Solanki}, S.~K., {Krivova}, N.~A., {Tagirov}, R.~V., \&
  {Schmutz}, W.~K. 2015, \aap, 581, A116

\bibitem[{{Shapiro} {et~al.}(2016){Shapiro}, {Solanki}, {Krivova}, {Yeo}, \&
  {Schmutz}}]{Shapiro2016}
{Shapiro}, A.~I., {Solanki}, S.~K., {Krivova}, N.~A., {Yeo}, K.~L., \&
  {Schmutz}, W.~K. 2016, {\aap}, 589, 14

\bibitem[{{Shapiro} {et~al.}(2013){Shapiro}, {Rozanov}, {Shapiro}, {Egorova},
  {Harder}, {Weber}, {Smith}, {Schmutz}, \& {Peter}}]{Shapiro2013}
{Shapiro}, A.~V., {Rozanov}, E.~V., {Shapiro}, A.~I., {et~al.} 2013, {\jgr},
  118, 3781

\bibitem[{{Skumanich}(1972)}]{Skumanich1972}
{Skumanich}, A. 1972, {\apj}, 171, 565

\bibitem[{{Soderblom}(2010)}]{Soderblom2010}
{Soderblom}, D.~R. 2010, {\araap}, 48, 581

\bibitem[{{Solanki} {et~al.}(2006){Solanki}, {Inhester}, \&
  {Sch{\"u}ssler}}]{Sami_B}
{Solanki}, S.~K., {Inhester}, B., \& {Sch{\"u}ssler}, M. 2006, Rep. Progr.
  Phys., 69, 563

\bibitem[{{Solanki} {et~al.}(2013){Solanki}, {Krivova}, \&
  {Haigh}}]{Solanki2013}
{Solanki}, S.~K., {Krivova}, N.~A., \& {Haigh}, J.~D. 2013, {\araap}, 51, 311

\bibitem[{{Tagirov} {et~al.}(2019){Tagirov}, {Shapiro}, {Krivova}, {Unruh},
  {Yeo}, \& {Solanki}}]{Rinat2019}
{Tagirov}, R.-V., {Shapiro}, A., {Krivova}, N., {et~al.} 2019, \aap, in press

\bibitem[{{Unruh} {et~al.}(1999){Unruh}, {Solanki}, \& {Fligge}}]{Unruh1999}
{Unruh}, Y.~C., {Solanki}, S.~K., \& {Fligge}, M. 1999, {\aap}, 345, 635

\bibitem[{{Wilson}(1978)}]{Wilson1978}
{Wilson}, O.~C. 1978, {\apj}, 226, 379

\bibitem[{{Yeo} {et~al.}(2014){Yeo}, {Krivova}, {Solanki}, \&
  {Glassmeier}}]{yeoetal2014}
{Yeo}, K.~L., {Krivova}, N.~A., {Solanki}, S.~K., \& {Glassmeier}, K.~H. 2014,
  \aap, 570, A85

\end{thebibliography}

\begin{appendix}
\section{Uncertainty of the irradiance variability}

The reconstruction of the irradiance variability presented in this study is based on the dependences of solar disk-area coverages by magnetic features on the chromospheric activity (see Eqs. 8--9). We have taken these dependences  from \cite{Shapiro2014} who first established them using  the observed solar coverages by magnetic features and chromospheric activity and then extrapolated the solar relationships to more active stars. The uncertainty of such an extrapolation is the main source of the uncertainty of our result.

\cite{Shapiro2014} employed several modified relationships of spot area coverages on the chromospheric activity to estimate the uncertainty of their results (see the detailed discussion in their Appendix A). Here we follow up on such an approach and consider following modified dependences:

\begin{equation}
    A_{\rm spots}^{\rm mod}(\rm S) = A_{\rm spots}(\rm S)\times (1+\alpha \: \frac{S - S_{\rm mean}^{\rm p}}{S_{\rm max} - S_{\rm mean}^{\rm p}}).
\end{equation}
Here $A_{\rm spots}^{\rm mod}(\rm S)$ is the modified disk-area coverage by spots, $A_{\rm spots}(\rm S)$ is the coverage obtained from Eq. (8), $S_{\rm mean}^{\rm p}$ is the mean value of the solar S-index over the solar activity cycle 23. Equation (A.1) implies that spot coverages obtained with Eq.~(8) are scaled with a factor $1+\alpha$ for stars with S-index value of $S_{\rm max}$. As in \cite{Shapiro2014}  we adopt $S_{\rm max}=0.5$ and recalculate the cyclic TSI variability with $\alpha = \pm 0.5$ (see Fig.~A.1). The green curves in Fig.~A.1 encompass the dependence shown in Fig.~3 and indicate the range corresponding to the uncertainty of the calculated amplitude of cyclic TSI variability.


\begin{figure}
\resizebox{\hsize}{!}{\includegraphics [angle=90] {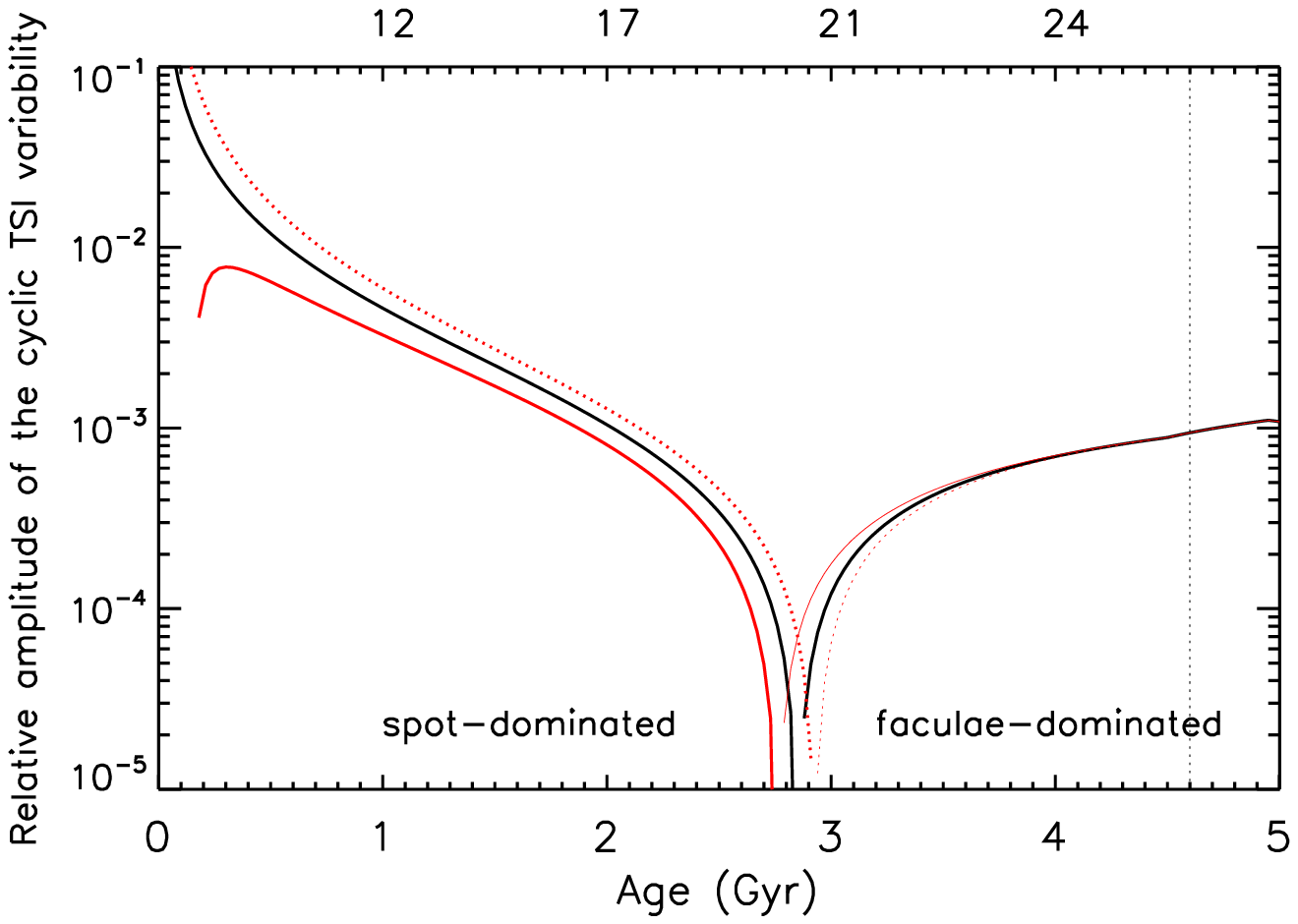}}
\caption{The same as Fig.~3 but without spot and facular components of the TSI variability. Shown are calculations with $\alpha=0$ (original  $A_{\rm spots}$ coverages given by Eq.~8, black), as well as $\alpha=0.5$ (increased spot coverages, dotted red) and $\alpha=-0.5$ (decreased spot coverages, solid red).}
\label{fig:prof_act}
\end{figure}


\end{appendix}

\end{document}